\documentclass[12pt]{article}
\newcommand{\ifour}{\int \frac{d^4 k}{(2 \pi)^4}}
\newcommand{\kslash}{\not \! k}
\newcommand{\dintst}{\int \int d^{4}x \; d^{4}y}
\newcommand{\qslash}{\not \! q}

\newcommand{\abp}{|\vec{p}|}
\newcommand{\dintsoT}{\int_0^{s_0(T)} \! \! \int_0^{s'_0(T)} \! \! \! 
\!ds\, ds'} 

\begin{document}
\mbox{ }
\rightline{UCT-TP-247/98}\\
\rightline{March 1998}\\
\vspace{3.5cm}
\begin{center}
{\Large \bf Pion-Nucleon Coupling at Finite Temperature}\\

\vspace{.5cm}

{\bf C. A. Dominguez$^{(a)}$, C. van Gend$^{(a)}$, and 
M. Loewe$^{(b)}$}\\[.5cm]
$^{(a)}$Institute of Theoretical Physics and Astrophysics\\
University of Cape Town, Rondebosch 7700, South Africa\\[.5cm]
$^{(b)}$Facultad de Fisica, Pontificia Universidad Catolica de Chile\\
Casilla 306, Santiago 22, Chile\\
\end{center}

\vspace{.5cm}

\begin{abstract}
\noindent
The pion nucleon vertex function at finite temperature is studied in
the framework of: (a) the thermal (linear) sigma model to leading (one-loop)
order, and (b) a thermal QCD-Finite Energy Sum Rule. Results from
both methods indicate
that the strength of the pion-nucleon coupling decreases with increasing
$T$, vanishing at a critical temperature. The associated mean-square
radius is a monotonically increasing function of $T$, diverging at the
critical temperature. This is interpreted as (analytical) evidence for
the quark-gluon deconfinement phase transition.
\end{abstract}
\newpage
\setlength{\baselineskip}{1.5\baselineskip}
\noindent
The temperature behaviour of hadronic Green's functions, and their 
associated parameters such as masses, widths, couplings, etc., has received
considerable attention lately, given its impact on the search for the 
quark-gluon plasma \cite{QGP}.
Two successful theoretical frameworks for these studies are the thermal
sigma model \cite{sigma1}-\cite{sigma3}, and QCD sum rules
\cite{QCDSR1}-\cite{QCDSR2}. The former technique provides information on
the $T$-dependence of pion and nucleon masses and widths associated, 
respectively, with the real and imaginary parts of their two-point Green's
functions. While these masses show no appreciable variation with temperature,
\cite{sigma1}-\cite{sigma2} their widths exhibit a dramatic increase with 
increasing $T$ \cite{sigma1},\cite{sigma3}. This result is in line 
with the expectation that
hadronic widths, interpreted as absorption coefficients in the thermal bath,
should diverge at some critical temperature \cite{gamma}. This provides
a signal or phenomenological order parameter for the quark-gluon 
deconfinement phase transition. Another such signal is the thermal behaviour
of hadronic couplings and form factors (three-point functions), which should
vanish at a critical temperature, where the associated  mean 
square radii should diverge. This has been explicitly confirmed for
the electromagnetic form factor of the pion \cite{Joel}, and for the
rho-pi-pi coupling \cite{Mirela}. In this note we study the $\pi NN$ vertex
function at finite temperature using the thermal (linear) sigma model,
as well as QCD sum rules. The purpose is to obtain additional confirming
(analytical) evidence for the deconfinement phase transition, as well
as information on this vertex function, which should be of use in hadron
gas models at finite temperature.

We begin with the (linear) sigma model, and consider the $\pi$NN vertex
\begin{equation}
\Gamma(q^2) = V(q^2) \overline{u}_f(p') \gamma_5 \tau_\alpha u_i(p) 
\end{equation}
where the nucleons are on-mass shell, and the pion has virtual mass
$q^{2} = (p' - p)^{2}$. 
The renormalization of the sigma model (at $T=0$) is discussed e.g. 
in \cite{Lee}, and the
renormalization of the $\pi$NN vertex (before the invention of dimensional
regularization and the $\bar{MS}$ scheme) may be found in \cite{M}. At 
finite temperature we
shall use the Dolan-Jackiw real time propagators \cite{DJ}, together with
the fact that thermal corrections do not induce any new kind of 
ultraviolet corrections. Hence, the thermal theory may be renormalized 
as at $T=0$. We have done this using dimensional regularization and the
$\bar{MS}$ scheme. To leading (one-loop) order, the relevant diagrams are
shown in Fig. 1 (a-e). At the kinematical point $q^{2}$=0, and in the
chiral limit, the expression for the irreducible vertex $V(q^{2})$ is
given by
\begin{equation} 
V(0) = g(1 + \beta(0) g^2 )
\end{equation}
where
\begin{eqnarray}
  \label{eq:3}
    \beta(0) & = & \frac{1}{16 \pi^2} \mbox{\huge[} -
  \frac{5}{3}\frac{M_\sigma^2}{M_N^2} + \frac{M_\sigma^2}{M_N^2}\left(3 -
    \frac{5}{6} \frac{M_\sigma^2}{M_N^2} \right)
  \ln\frac{M_\sigma^2}{M_N^2} \nonumber \\
  & &   \! \! \! \! \! \!+ \frac{1}{3}(5 M_\sigma^2 - 8 M_N^2) \frac{M_
  \sigma}{M_N^4} \sqrt{4 M_N^2
    - M_\sigma^2} \;\;\arctan\frac{\sqrt{4 M_N^2 - M_\sigma^2}}{M_\sigma}
    \mbox{\huge ]}
\end{eqnarray}

with $M_{N}$ and $M_{\sigma}$ being the nucleon  and sigma
meson masses, respectively. Equation (2) may be regarded as an expression 
for the effective coupling constant $g_{\pi NN}$ in the chiral limit. It is
equal to $g$ if $\beta(0)$ vanishes. This happens for $M_{\sigma} \simeq
1300 \;MeV$; in fact, $\beta(0)$ is small and negative if $M_{\sigma}$ is
bigger than this value.
One should recall that in the linear sigma model $g_{A} =1$, so
that the Goldberger-Treiman relation (GTR) \cite{GTR} 
becomes: $M_{N} = g f_{\pi}$.
Using the chiral symmetry limit values \cite{GLPR}: $M_{N} \simeq 800 \;MeV$,
and $f_{\pi} \simeq 80 \;MeV$, one finds: $g \simeq$ 10, not far from the
experimental value $g_{\pi NN} \simeq$ 13. In any case, here we are only 
interested in the temperature behaviour of the {\it ratio}
$V(q^{2},T)/V(q^{2},0)$; particularly in the possibility that this ratio
vanishes at a critical temperature, and that the  mean square radius
diverges there. This will turn out to be largely independent of the 
particular value assumed
by $g$ or $g_{\pi NN}$ and $\beta(0)$ (or equivalently $M_{\sigma}$),
although the specific value of the critical temperature does depend 
on the latter.\\

Turning to the temperature corrections to the graphs shown in Fig. 1,
we need only consider Fig. 1(b) and Fig. 1(d) (identical to Fig. 1(e)), 
as Fig. 1(c) is Boltzmann suppressed on account of $M_{N} \simeq M_{\sigma}
>> \mu_{\pi}$. The thermal correction to the graph Fig. 1(b) is given by

\begin{equation}
  \label{eq:4}
  \Gamma(q^2,T) = -g^3 \overline{u}_f(p') \gamma_5 \tau_\alpha u_i(p) \int
  \frac{d^4 k}{(2 \pi)^4} \frac{k^2 \, n_B(k_0) \delta(k^2 - 
  \mu_\pi^2)}{[(p'-k)^2 - M_N^2][(p-k)^2 - M_N^2]}
\end{equation}

where $n_{B}$ is the Bose thermal factor: $n_{B}(z) = (e ^{z/T} - 1)^{-1}$.
Hence, this contribution vanishes in the chiral limit. That of Fig. 1(d)
is found to be

\begin{equation}
  \label{eq:5}
  \Gamma(q^2,T) = g^3 \frac{(M_\sigma^2 - \mu_\pi^2)^{2}}{2 f_{\pi}^{2}M_N}
  \overline{u}_f (p')\gamma_5 \tau_\alpha \ifour \frac{(2 m_N - \kslash)2
    \pi \delta(k^2) n_B(k_0)}{[(k+q)^2 - M_\sigma^2][(p-k)^2 - 
      M_N^2]} u_i (p)
\end{equation}

We choose for convenience a Lorentz frame in which the incoming nucleon
is at rest with respect to the heat bath ($(\vec{p} = 0$). We have 
checked that the final results are largely independent of the choice
of frame.
The thermal effective $\pi NN$ coupling in the chiral limit, and at 
$q^{2}$=0, may then be written as

\begin{equation}
  \label{eq:6}
    \frac{V(0,T)}{ V(0,0)}= 1 -\frac{g^{2}T^{2}}{12 M_N^2(1+g^{2}
    \beta(0))}
\end{equation}
 
where the GTR has been used. Notice that an extrapolation in temperature 
of this result implies a critical temperature
\begin{equation}
  \label{eq:7}
    T_d = \sqrt{12(1 + \beta(0) g^{2})} \, M_N /g 
\end{equation}
which depends on the value of the sigma-meson mass through $\beta(0)$;
numerically, $T_{d} \simeq 150 - 300\; MeV$ if $M_{\sigma} 
\simeq 1300 - 1600\;
MeV$. One should not assign too much importance to the specific numerical
values of this critical temperature and/or the sigma-meson mass,
to wit. First,
the relations among parameters in the sigma model are valid at the 25-30\%
level; e.g. $g_{A}=1$ instead of the experimental value $g_{A}$ = 1.26, 
and $g_{\pi NN} \simeq$ 10 (from the GTR), as opposed to the experimental 
value $g_{\pi NN} \simeq$ 13, etc.. Second, Eq.(6) and the $T^{2}$ dependence
is a consequence of the one-loop approximation. Higher loop corrections
will induce higher order (in $T$) terms which will alter the numerical value
of the critical temperature. These will be suppressed, though, by inverse
powers of the nucleon/sigma-meson masses. A similar situation arises in
chiral perturbation theory and the $T$-dependence of the pion decay constant
$f_{\pi}(T)$. To order $T^{2}$, $f_{\pi}(T)$ vanishes at $T_{c} \simeq
240 \;MeV$ \cite{CHPT}, while higher order corrections bring down 
this value considerably. What we find important here, is that the $\pi NN$
coupling at leading order in $T$ decreases with increasing temperature.

Next, we consider the mean-square radius associated with 
$V(q^{2},T)$, and defined as

\begin{equation}
  \label{eq:8}
    \frac{< r^{2}_{\pi NN}> _{T}} 
    {< r^{2} _{\pi NN} > _{0}} 
    = \frac{V(0,0)}{V(0,T)} \mbox{\huge[} \left. \frac{\partial
        V(q^2,T)}{\partial q^2} 
        \left / \frac{\partial V(q^{2},0)}    
        {\partial q^{2}} \right . \mbox{\huge]}
        \right|_{q^2 = 0}
\end{equation}

We have calculated this ratio numerically, the result being shown in 
Figure 2. An extrapolation in temperature indicates quite clearly
the divergence of this  mean square radius. This may be interpreted
as a signal for deconfinement, to the extent that the {\it size} of
the nucleon, as probed by a pion, increases with increasing temperature,
becoming infinite at $T=T_{d}$.\\ 

We study next the same vertex function, but in the
framework of QCD Finite Energy Sum Rules (FESR). This will provide
important independent support to the above result, especially since
the QCD sum rule technique, unlike the sigma model, does not entail
any expansion in powers of the temperature. To this end, we begin the
analysis at zero temperature and introduce the three-point function

\begin{equation}
  \label{eq:9}
    \Pi(p,p',q) = i^2 \dintst \langle 0 \left| T\left(\eta(x) J_5(y)
      \overline{\eta}(0) \right) \right| 0 \rangle e^{i (p'\cdot x - q
    \cdot y)}
\end{equation}

where the nucleon and pion interpolating currents, $\eta(x)$ and $J_{5}(x)$,
respectively, are chosen as
\begin{equation}
  \label{eq:10}
  \eta(x) = \epsilon_{abc} \left[ u^a(x) C \gamma_\mu u^b(x) \right]
  \gamma^\mu \gamma_5 d^c(x)
\end{equation}
\begin{equation}
  \label{eq:11}
    J_5(x) = i \left[\overline{u}(x) \gamma_5 u(x) - \overline{d}(x)
    \gamma_5 d(x) \right]
\end{equation}

where $C$ is the charge conjugation operator.
The couplings of these currents to the nucleon and the pion are defined as
\begin{equation}
  \label{eq:12}
  \langle 0|\eta(0)|N(p,s)\rangle = \lambda_N u(p,s)
\end{equation}
\begin{equation}
  \label{eq:13}
    \langle N(p_2,s_2) | J_5(0) | N(p_1,s_1) \rangle =
  \overline{u}(p_2,s_2) \gamma_5 g_{P}(q^{2})
    u(p_1,s_1)
\end{equation}
where
\begin{equation}
  \label{eq:14}
  g_P(q^{2}) =  \frac{f_{\pi}\mu_\pi^2}{m_q}
  \frac{g_{\pi NN}}{q^{2}-\mu_{\pi}^{2}}
\end{equation}
and where $m_{q}$ is the average of the up and down quark masses. 
In our normalization, the pion decay constant is $f_{\pi} \simeq 93 \;MeV$.
The hadronic representation of the imaginary part of the vertex function
Eq. (9) is obtained by inserting a complete set of hadronic states. After
summing over spins, and making the standard nucleon-pole approximation
(thus including in the continuum all the radial excitations of the nucleon)
one obtains

\begin{eqnarray}
  \label{eq:15}
  Im \Pi(s,s',q^2)|_{\mbox{HAD}} & = & g_P \lambda_N^2 M_N 
  \left(i \gamma_5 \qslash \right) \pi^2 \delta(s - M_N^2) \delta(s' -
  M_N^2)   \nonumber \\ 
  & &   + \theta(s - s_0) \theta(s' - s'_0) Im
  \Pi(s,s',q^2)|_{\mbox{QCD}}
\end{eqnarray}

Since we are interested in the pion-nucleon coupling in the vicinity
of $q^{2}$=0, we can safely neglect any $q^{2}$ dependence in $g_{\pi NN}$.
This dependence would arise from the contribution of the radial excitations
of the pion, $\pi'$(1300) etc., which in the chiral limit is a correction
of order $q^{2}/M_{\pi'}^{2}$.
As usual, the hadronic continuum with thresholds $s_{0}$ and $s'_{0}$ is
modelled by the QCD spectral function. The leading order diagrams 
needed to compute the latter are shown in Fig. 3. In the chiral limit,
the relevant structure to be sought is proportional to 
$\not \!\! q q^{2}$.
It turns out that the diagram Fig. 3a does not have this behaviour,
while that of Fig. 3b (plus all other related diagrams) gives

\begin{equation}
  \label{eq:16}
  Im \Pi(s,s',q^2)|_{\mbox{QCD}} = \frac{\langle \overline{q}
    q\rangle}{2 \pi} \frac{i \gamma_5 \qslash}{q^2} (s + s')
\end{equation}

where $<\bar{u} u> \simeq <\bar{d} d> = <\bar{q} q>$ has been used.
By means of Cauchy theorem, and assuming global (quark-hadron) duality,
one obtains the lowest dimensional FESR

\begin{equation}
  \label{eq:17}
  g_{\pi NN} = \frac{f_{\pi}}{8 \pi^{3}}
  \frac{s_{0} s'_{0} (s_{0}+s'_{0})}
  {\lambda_{N}^{2} M_{N}}
\end{equation}

where use has been made of the Gell-Mann, Oakes and Renner relation
\cite{GMOR}
\begin{equation}
  \label{eq:18}
    f_\pi^2 m_\pi^2 = - 2 m_q \langle \overline{q} q \rangle
\end{equation}
Since the dispersion in $p^{2} = s$ and $p'^{2} = s'$ refers to the two
nucleonic legs, it is reasonable to assume $s_{0} \simeq s'_{0}$.
An analysis of the two-point function involving the nucleonic current
$\eta(x)$ \cite{NUCL} in the framework of QCD FESR yields

\begin{equation}
  \label{eq:19}
    \lambda_N^2  =  \frac{s_0^3}{192 \pi^4}\;\;\;\;\; \;\;\;\;\;\;
\lambda_N^2 M_N  =  - \frac{\langle \overline{q} q\rangle}{8 \pi^2} 
  s_0^2
\end{equation}

which determines the nucleon mass in terms of $s_{0}$. Conversely, using
$M_{N}$ and $<\bar{q} q>$ as input, Eq.(19) fixes $\lambda_{N}$ and $s_{0}$.
In this fashion, Eq.(17) becomes: $g_{\pi NN} = 48 \pi f_{\pi} / M_{N}
\simeq 15$, not far from the experimental value $g_{\pi NN} \simeq 13$.
This level of agreement is more than enough for our purpose here, which is
to determine the temperature behaviour of the $\pi NN$ coupling, i.e. the
ratio $g_{\pi NN}$(T) $g_{\pi NN}$(0). To achieve this, we have calculated 
the thermal corrections to the QCD spectral function with the result
\begin{equation}
  \label{eq:20}
    Im \Pi(p,p',q) = \frac{1}{4 \pi} \langle \overline {q} q \rangle
  \frac{i \gamma_5 \qslash}{q^2} \left\{ p^2 f(p,T) + p'^2 f(p',T) \right\}
\end{equation}
where

\begin{equation}
  \label{eq:21}
  f(p,T) = \int_{-1}^1 dx \left[ 1 -  n_F\left(\frac{|p_0 - \abp x|}{2}\right) - n_F\left(\frac{|p_0 +
          \abp x|}{2}\right) \right]
\end{equation}

and a similar expression for $f(p',T)$. The FESR in this case becomes

\begin{equation}
  \label{eq:22}
  g_{\pi NN}(T)=\frac{1}{8 \pi^{3}}
  \frac{f_{\pi}(T)}{M_{N}(T) \lambda_{N}^{2}(T)}
    \dintsoT  \left[ s
    f(p,T) + s' f(p',T) \right] 
\end{equation}

where use has been made of the thermal Gell-Mann, Oakes and Renner
relation, which has been recently shown to be valid over a wide range of
temperatures \cite{GMORT}. The temperature behaviour of $f_{\pi}$,
valid up to the critical temperature, has been obtained in \cite{BAR1}.
The function $s_{0}(T)$ has been determined from a FESR for the two-point
function involving the axial-vector current \cite{s0} - \cite{BAR2};
it scales as: $s_{0}(T) /s_{0}(0) \simeq ( f_{\pi}(T)/f_{\pi}(0))^{2}$.
The thermal nucleonic coupling $\lambda_{N}(T)$ has been determined from 
a QCD-FESR in the nucleon channel, with the result \cite{NUCL}
\begin{equation}
  \label{eq:23}
    \lambda_N^2(T) = \lambda_N^2(0) \left( \frac{s_0(T)}{s_0(0)}
  \right)^3 \left[1 + G_a(T) \right]
\end{equation}
where

\begin{eqnarray}
  \label{eq:24}
    G_a(T) & = & \frac{576}{\left(s_0(T)\right)^3}
  \int_0^{\sqrt{s_0(T)}} \! \! d\omega \int_0^{\omega/2} \! \! dx
  \int_{\omega/2-x}^{\omega/2} \! \! dy\  x (\omega - 2 x) \left\{
    -n_F(x) \right. \nonumber \\ 
& & 
  \left. - n_F(y) + n_F(x) n_F(y) + n_F(\omega -x-y) \left[ n_F(x) + n_F(y) -
      1\right] \right\}.
\end{eqnarray}

and $n_{F}(z) = (1 + e^{-z/T})^{-1}$ is the Fermi factor. With all this
information one can then solve Eq.(22), after choosing a particular Lorentz
frame. Our choice is the rest frame of the incoming nucleon, i.e. 
$\vec{p} = 0$ and $p_{0}= \sqrt{s}$, however, the final results are quite
insensitive to the choice of frame. The result for the ratio 
$g_{\pi NN} (T)/ g_{\pi NN} (0)$ as a function of $T/T_{d}$ is shown 
in Fig.4.
One can  clearly appreciate the vanishing of this coupling at the critical
temperature. This is the temperature at which $f_{\pi}(T)$ vanishes,
i.e. the critical temperature for the chiral-symmetry restoration
phase transition, which is basically the same temperature at which
$s_{0}(T)$ vanishes, i.e. the critical temperature for the quark-gluon
deconfinement phase transition \cite{s0} - \cite{BAR2}.\\

Finally, the mean-square radius associated to the pion-nucleon vertex
\begin{equation}
  \label{eq:25}
  \langle r_{\pi NN}^2 \rangle _{T} = 6 \frac{\partial}{\partial q^2}
  \ln{g_{\pi NN}(q^2,T)}|_{q^2 = 0}
\end{equation}
can be easily calculated, with the result
\begin{eqnarray}
  \label{eq:26}
    \langle r_{\pi NN}^2 \rangle _{T} & = & \left \{ \dintsoT \!
      \int_{-1}^1 \! \! dx \left[s \left(1 - 2 n_F (\sqrt{s}/2)
        \right) + s' \left(1 - 2 n_F(z) \right) \right] \right\}^{-1}
    \nonumber \\
  & & \times\frac{6}{2 T} \dintsoT \frac{1}{\sqrt{s}} \int_{-1}^1
  \! \! dx \, n_F^2(z) e^{z/T} \left[ 1 + x \frac{s + s'}{|s - s'|} \right]
\end{eqnarray}
where
\begin{equation}
  \label{eq:27}
    z = \frac{s + s' + |s - s'| x}{4 \sqrt{s}}
\end{equation}

and the rest frame of the incoming nucleon has been used. A numerical 
evaluation of Eq.(26) gives the result shown in Fig. 5. Notice that since
we have made the pion-pole approximation, $g_{\pi NN}$ at $T=0$ is 
independent of $q^{2}$. The mean-square radius is non-vanishing only
at finite temperature, where a $q^{2}$ dependence appears through the
Fermi factors.\\

In summary, the vanishing of the pion-nucleon coupling, and the divergence
of the associated mean-square radius, at a critical temperature
has been shown to follow from the thermal linear sigma model at leading
(one-loop) order, as well as from a thermal QCD-FESR. This may be viewed
as (analytical) evidence supporting the existence of the 
quark-gluon deconfinement phase transition. As the critical temperature
is approached, the strength of the coupling of pions to nucleons 
is quenched, and at the same time, the size of the nucleon as probed
by the pion gets bigger. The qualitative
agreement between the two methods lends further support to the extension
of the QCD sum rule program to finite temperature. It should be noticed that
potential non-diagonal vacuum condensates \cite{H} do not enter our FESR
because of their (higher) dimensionality. We have emphasized many times in
the past that QCD-FESR are far better than e.g. QCD-Laplace transform
sum rules, to the extent that the lowest dimensional thermal FESR do 
not involve (unknown) non-diagonal vacuuum condensates.\\[.3cm]

{\bf Acknowledgements}\\
This work  has been supported in part by the Foundation
for Research Development (South Africa), and by Fondecyt (Chile) under
grant No. 1950797.

\begin{center}
{\bf Figure Captions}
\end{center}
Figure 1. Leading (one-loop) diagrams contributing to $g_{\pi NN}$ in
the linear sigma model.\\
Figure 2. Thermal behaviour of the $\pi NN$ mean square radius, Eq.(8), 
in the linear sigma model.\\
Figure 3. Leading order QCD diagrams entering the  determination 
of the spectral function relevant to $g_{\pi NN}$.\\
Figure 4. Thermal behaviour of the $\pi NN$ coupling determined from the 
QCD-FESR Eq.(22).\\
Figure 5. Thermal behaviour of the $\pi NN$ mean square radius, Eq.(26), 
according to the QCD-FESR.\\
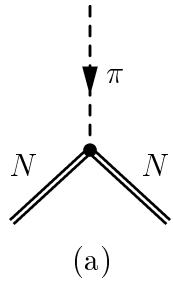
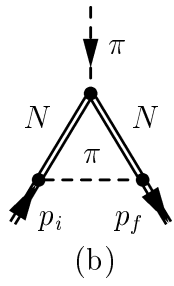
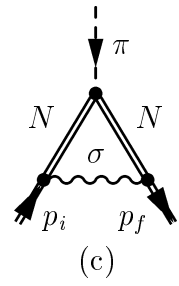
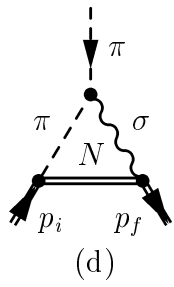
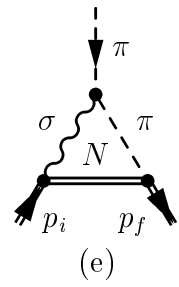
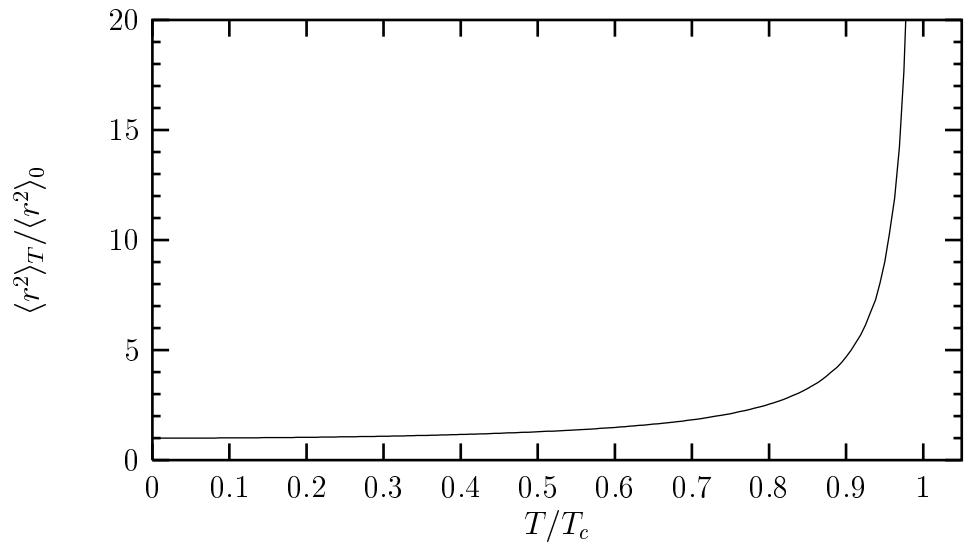
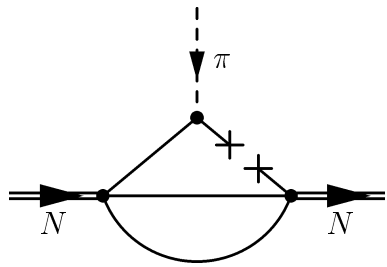
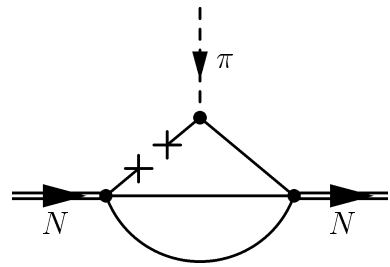
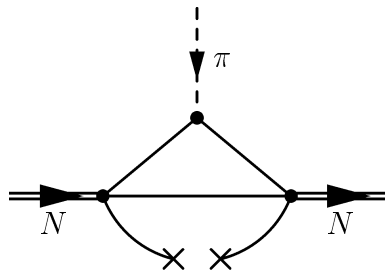
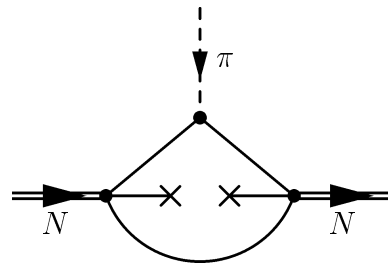
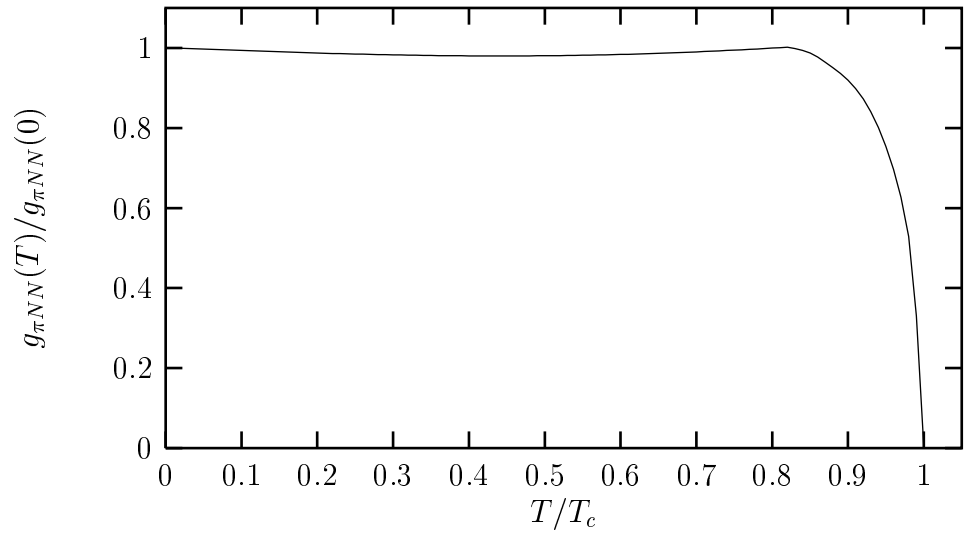
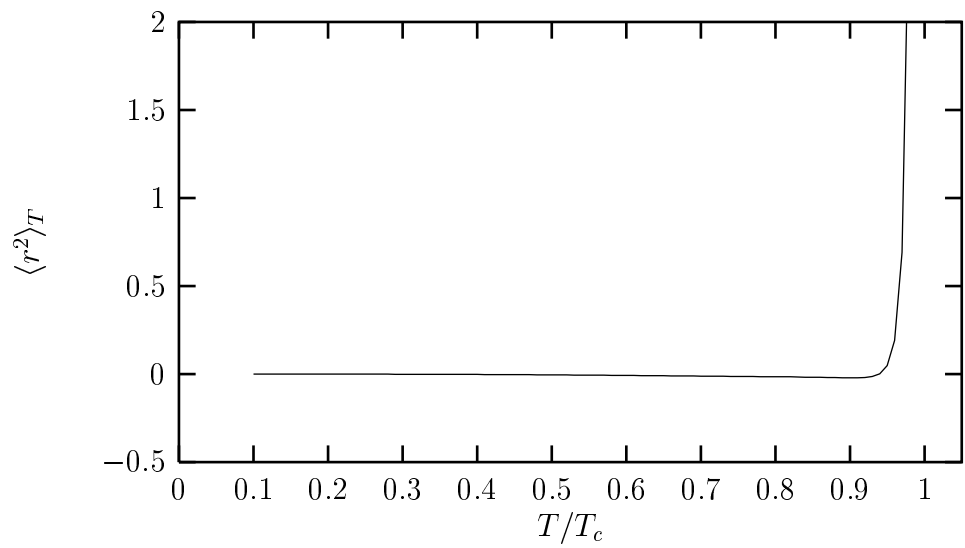
\end{document}